\documentclass[aps,twocolumn,showpacs]{revtex4}
\usepackage{amsmath}
\usepackage{amssymb}
\usepackage{amsthm}
\usepackage{amsfonts}
\usepackage{enumerate}
\usepackage{latexsym}
\usepackage{graphicx}

\newcommand{\beq}{\begin{equation}}
\newcommand{\eneq}{\end{equation}}
\input{epsf}

\begin{document}

\title{Classical spins in topological insulators}
\author{Qin Liu$^1$ and Tianxing Ma$^2$}
\affiliation{$^{1}$Department of
Physics, Fudan University, Shanghai 200433, China\\
 $^{2}$ Max-Planck-Institut f\"ur Physik Komplexer Systeme, N\"othnitzer Strasse 38, 01187 Dresden, Germany}
\date{\today}

\begin{abstract}
Following the recent theoretical proposal and experiment on quantum
spin Hall effect in HgTe/CdTe quantum wells, we consider a single
magnetic impurity localized in the bulk of the system, which we
treat as a classical spin. It is shown that there are always
localized excited states in the bulk energy gap for arbitrarily
strong impurity strength in inverted region, while the localized
excited states vanish for very strong impurity strength in normal
region. Similar conclusion also applies to three-dimensional
topological insulators. This distinct difference serves as another
novel criterion for the conventional and topological insulating
phases when the time-reversal symmetry is broken, and can be easily
experimentally observed through the STM and/or ARPES experiments.
\end{abstract}


\maketitle
\tolerance 10000

The quantum spin Hall (QSH) effect is a state of matter with
topological properties distinct from those of conventional
insulators. \cite{Kane,Bernevig1,Bernevig2006D} The first proposal
of experimental realization of this effect is given in the work by
Bernevig {\it et al.} \cite{Bernevig2006D} where they consider the
HgTe/CdTe semiconductor quantum wells (QWs), and show that when the
thickness of the QW is varied, the electronic states change from a
normal to an ¡°inverted¡± type at a critical thickness $d_c$. This
transition is a topological quantum phase transition between a
conventional insulating phase and a phase exhibiting the QSH effect
with a single pair of helical edge states. This phase transition can
be understood by the relativistic Dirac model in (2+1) dimensions,
which mimic the electronic states near the $\Gamma$ point. At the
quantum phase transition point, $d$=$d_c$, the mass term in the
Dirac equation changes sign, leading to two distinct $U(1)$-spin and
$Z_2$ topological numbers on either side of the transition.
Recently, the QSH phase in HgTe/CdTe QWs has been observed in the
transport experiments, \cite{Konig2007} which confirms Bernevig {\it
et al.}'s theoretical predictions.\cite{Bernevig2006D}

Following this pioneer work, there emerge various discussions on the
novel properties of the topological insulating phase in both two-
and three-dimensional (2-, 3-D) systems,
\cite{Roy2006,Moore2007,Fu2007,Fu2008} however, most of these are
considered within the framework of the preservation of the
time-reversal symmetry (TRS), among which, we notice that two of
them show that the novel properties of this topological system can
also be manifested by breaking the TRS on the surface through the
so-called topological magneto-electric effect \cite{Qi2008a} or
local charge and spin density of states. \cite{liu2008a} In the
meanwhile, we notice that the system with Mn doped impurities in the
bulk of the HgTe QWs has been discussed in Ref.\cite{lcx2008}, where
by breaking the TRS in the bulk, the quantum anomalous Hall effect
is realized. On the other hand, it is well known that a single
magnetic impurity in a superconductor breaks the TRS and induces low
energy bound states in the superconducting gap. \cite{Yu,Shiba1968}

Motivated along this line, we discuss the presence of a single
magnetic impurity located in the 2D bulk of the HgTe/CdTe QWs, which
we treat as a classical spin in both normal and inverted regimes.
Similar to the discussions in BCS superconductors by Shiba in 1968
\cite{Shiba1968}, we show that in the inverted regime of the
HgTe/CdTe QWs, there are always localized excited states (LES) in
the bulk energy gap for arbitrarily strong impurity strength, while
in normal regime, the LES vanish into the bulk for very strong
impurity strength. This distinct difference of the response to the
single magnetic impurity in bulk serves as another novel criteria
for the conventional and topological insulating phases when the TRS
is broken, and can be experimentally observed through the STM and/or
ARPES measurements.

The starting point of this paper is the effective four-band
model\cite{Bernevig2006D}
$H_{0}({\vec{k}})=\varepsilon_k\sigma_0\tau_0+\mathcal{M}_{k}\sigma_0\tau_3+Ak_{x}\sigma
_3\tau_1+Ak_{y}\sigma _0\tau_2$ in HgTe/CdTe QWs
around the $\Gamma$ point in the basis of $\left\vert E1,+\right\rangle ,$ $%
\left\vert H1,+\right\rangle ,$ $\left\vert E1,-\right\rangle ,$
$\left\vert H1,-\right\rangle$, plus that of a short-range single
magnetic impurity located at the origin. The exchange interaction in
Mn doped HgTe QWs has been discussed in several literatures,
\cite{Novik2005,lcx2008} where it is established that the $s$-band
and $p$-band electrons have different $sp$-$d$ exchange coupling
strength. To focus on the physical picture, we first consider the
isotropic case where $J^s=J^p=J$, then the full Hamiltonian takes
the form
\begin{equation}
H=\sum_{k}c_{k}^{\dag }H_{0}(\vec{k}) c_{k}
+\frac{J}{2}\sum_{kk^{\prime }}c_{k}^{\dag }(\vec{S}\cdot
\vec{\sigma }) \tau_0c_{k^{\prime }} \label{Hamil}
\end{equation}
Here $\vec{S}$ is the spin vector of the magnetic impurity, and the
Pauli matrix $\sigma_i$'s act on spin space while $\tau_i$'s act on
the two electric subbands space, $\sigma_0$ and $\tau_0$ are both 2
by 2 unit matrices. We will show later that our result is robust for
a general form of the exchange coupling. The full Green's function
(GF) of Hamiltonian (\ref{Hamil}) is obtained through the equation
of motion formulation \cite{Shiba1968} as
\begin{eqnarray}
G_{kk^{\prime}}(\omega)=G^{0}_{k}( \omega)
\delta_{kk^{\prime}}+G^{0}_{k}(\omega) t(\omega)
G^{0}_{k^{\prime}}(\omega) \label{GF}
\end{eqnarray} where the
t-matrix takes the form
\begin{equation}
t\left( \omega \right)=\frac{\left( \frac{JS}{2}\right)
^{2}F\left( \omega \right) +\frac{J}{2}\left( \vec{S}\cdot \vec{\sigma}%
\right) \tau _{0}}{1-\left[ \frac{JS}{2}F\left( \omega \right)
\right] ^{2}} \label{tmatrix}
\end{equation}
with $S^{2}=S_{x}^{2}+S_{y}^{2}+S_{z}^{2}$. In the above, we have
introduced a $F$-function as
\begin{equation}
F(\omega)=\frac{1}{N}\sum_{k}G^0_k( \omega)=\text{diag}(F_e\; F_h \;
F_e\; F_h) \label{Fmatrix}
\end{equation}
 where the diagonal elements are
\begin{equation}
F_{e(h)}(\omega)=\frac{1}{N}\sum_{k}\frac{\omega -\varepsilon
_{k}+(-)\mathcal{M}_k}{D_k} \label{Ffunction}
\end{equation}
In the tight-binding model, $\varepsilon _{k}=C-2D(2-\cos k_{x}-\cos
k_{y})$,
$\mathcal{M}%
_{k}=M-2B(2-\cos k_{x}-\cos k_{y})$ and $D_{k}=\left(\omega
-\varepsilon _{k}-\mathcal{M}_{k}\right)\left( \omega -\varepsilon
_{k}+\mathcal{M}_{k}\right)-A^{2}(\sin ^{2}k_{x}+\sin ^{2}k_{y})$,
where $M,\; A,\; B,\;C\;,D$ are material parameters introduced in
the effective four-band model in Ref. \cite{Bernevig2006D}. Then the
eigen-energies for the excited states are obtained by finding the
poles of the GF in the bulk energy gap as
\begin{eqnarray}
\frac{JS}{2}F_{e(h)}\left( \omega \right) =\pm 1
\label{resonant}
\end{eqnarray}
which is consisted of four equations.
\begin{figure}[tbp]
\begin{center}
\includegraphics[width=2.6in] {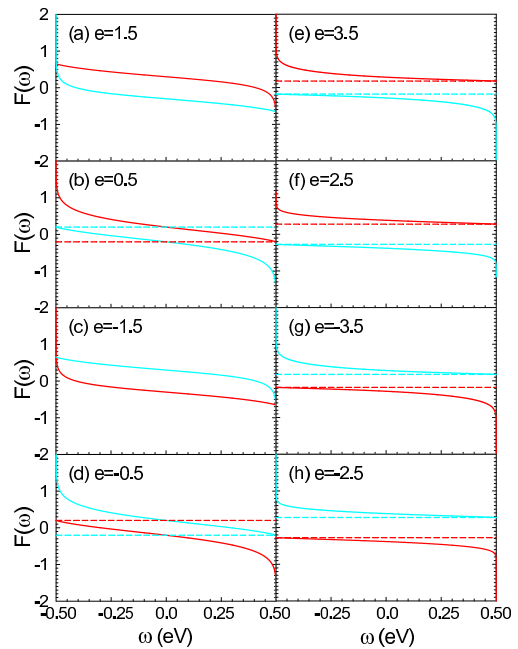}
\end{center}
\caption{(Color online) $F$-function versus $\omega $ in the bulk
energy gap with several values of $e$ and $c^2=1$. The topological
nontrivial regime with $|e|<2$ is shown on the left column, while
the topological trivial regime with $|e|>2$ is shown on the right
column of the figure. $F_{e}(e,\omega )$ is plotted as red lines
while $F_{h}(e,\omega )$ is plotted as cyan lines. The dotted lines
are guidelines for eyes to indicate the boundary values of the
$F$-functions.} \label{F2D}
\end{figure}

\begin{figure}[tbp]
\begin{center}
\includegraphics[width=2.2in] {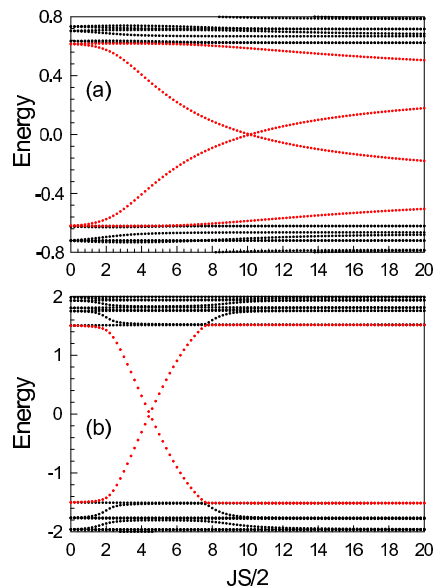}
\end{center}
\caption{(Color online) Energy spectrum obtained by direct
diagonalization of Hamiltonian (\ref{Hamil}) with $\epsilon_k=0$ as
a function of impurity strength in both (a) inverted and (b) normal
regimes. The localized states are shown in red.} \label{diag2D}
\end{figure}
To characterize the momentum integration in both normal and inverted
regimes, we notice that each diagonal block in $H_0(\vec{k})$
describes a quantum anomalous Hall system, \cite{Qi2006B} with the
two forming a time-reversal conjugate pair. Using the result
obtained by Qi {\it et al.},\cite{Qi2006B} the topological behavior
of this system is totally governed by two key parameters. In our
case the correspondences are $e=-\frac{M}{2B}+2$, which is related
to the mass term in the (2+1)-D Dirac model, and $c=-\frac{2B}{A}$
which determines the sign of the Chern number. Therefore in the
inverted regime we have $|e|<2$ corresponding to $d>d_c$, otherwise
it is topological trivial corresponding to a normal insulating
phase. Furthermore, we set $\varepsilon _{k}=0$ without loss of
generality to obtain a particle-hole symmetric system, as we know
that the quadratic kinetic term has no contribution to the topology
of this system. \cite{Qi2006B}

We therefore rewrite Eq.(\ref{Ffunction}) in terms of $e$ and $c$,
the result of which with $\omega$ in the bulk energy gap at several
values of $e$-parameter is shown in Fig.\ref{F2D}, where the
topological nontrival regime with $|e|<2$ is given on the left
column, while the trivial regime is plotted in the right column with
$|e|>2$. In each panel, the contribution from the electron subband,
$F_e(\omega)$, is shown in red and that from the hole subband,
$F_h(\omega)$, is shown in cyan. It is clear to see that the
resonant condition, Eq.(\ref{resonant}), is always satisfied for any
given impurity strength $J$ in the inverted regime, and there are
four LES in the bulk gap, two come from the electron subband and two
from the hole subband. Moreover, the stronger the impurity strength
is, the nearer the LES are to the middle of the bulk gap, and no
matter how strong the impurity strength is, there are always LES in
the bulk gap which appear as peaks in the density of states (DOS).

In contrast, for the normal regime, we see that the resonant
condition is {\it not} always satisfied for any impurity strength.
For weak impurity strength, there could exist two LES near the gap
edge, however, as we increase the impurity strength, these LES merge
quickly into the bulk and vanish finally. That is, the peaks in the
DOS in the bulk gap are not stable against the strong impurity
strength for this case. This distinct difference of the response to
the single magnetic impurity in bulk serves as another novel
criteria for the conventional and topological insulating phases when
the TRS is broken.
\begin{figure}[tbp]
\begin{center}
\includegraphics[width=2.4in] {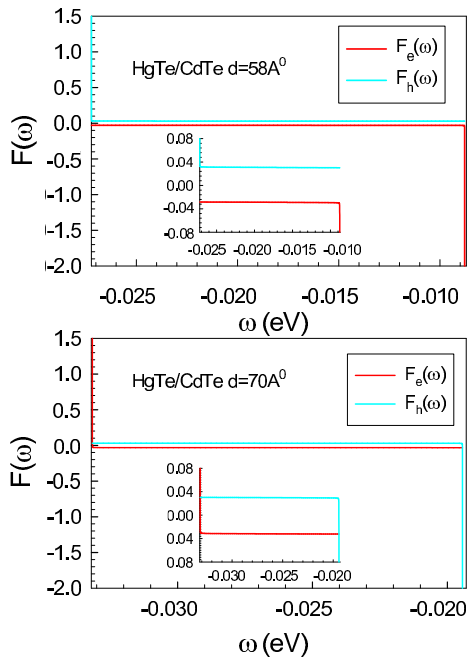}
\end{center}
\caption{(Color online) $F$-function versus $\protect\omega $ with
$\protect\omega $ in the bulk energy gap. The $F$-functions are
calculated numerically from Eq. (\ref{Ffunction}) with the
parameters taken from Ref. \cite{Bernevig2006D} for
HgTe/CdTe QW at $d=58{\mathring{A}}$ and $d=70{\mathring{A}}$. $%
F_{e}(\protect\omega )$ are shown as red lines while
$F_{h}(\protect\omega )$ are shown as cyan lines. Inset: the
enlarged structures for the nontrivial regime.} \label{HgTe}
\end{figure}
Note that this result is robust to the explicit form of the exchange
interaction. For the Mn doped HgTe QWs \cite{Novik2005,lcx2008}, we
consider a form of exchange interaction,
$J^{s(p)}_zS_z\sigma_z+J^{s(p)}_{\shortparallel}(S_x\sigma_x+S_y\sigma_y)$,
in electron and hole bands separately. It turns out that the only
difference with the isotropic model is to replace $JS$ in
Eq.(\ref{resonant}) by
$\sqrt{(J_z^{s,p})^2S^2_z+(J_{\shortparallel}^{s,p})^2S^2_{\shortparallel}}$,
and in the inverted region there are still four persistent LES in
the gap, which never merge into the bulk.

To justify the above results obtained from t-matrix method, we
directly diagonalize the Hamiltonian (\ref{Hamil}) on a square
lattice in both inverted and normal regimes by taking
$e=0.5\;\text{and}\;3.5$ for example. The obtained energy spectrum
is plotted versus impurity strength in Fig.\ref{diag2D}. The four
persistent LES (red lines) in the bulk energy gap are clearly shown
for the nontrivial case where we see that they approach the middle
of the gap as the increasing of $J$ and do not vanish. While in the
trivial regime, there are only two LES for small $J$ and merge into
the bulk for large $J$. This result is in perfect agreement with the
above analysis through GF discussions.

By using the real parameters of the HgTe/CdTe QWs system given in
Ref. \cite{Bernevig2006D}, we plot the $F$-function at the quantum
well width $d=58\;{\mathring{A}}$ and $d=70\;{\mathring{A}}$, which
are shown in Fig.\ref{HgTe}. We see that for $d>d_c$ where the QSH
effect is predicted, there are always LES in the bulk energy gap for
arbitrarily strong impurity strength; while for $d<d_c$, which is a
normal insulator, the LES vanish for strong impurity strength.

\begin{figure}[tbp]
\begin{center}
\includegraphics[width=2.8in] {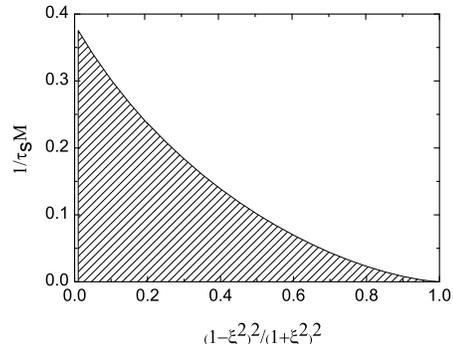}
\end{center}
\caption{Effective impurity concentration $\frac{1}{\tau_s M}$ as a
function of effective impurity strength $(1-\xi^2)^2/(1+\xi^2)^2$.
The shaded area indicates the range of the impurity concentration
where the impurity band and the continuum are separated.} \label{ni}
\end{figure}

Considering that in real materials there is always a finite
concentration of magnetic impurities, under which the localized
excited states grow into impurity band, we estimate the critical
concentration of magnetic impurities $\frac{1}{\tau_s M}$ in the
topological nontrivial regime as a function of impurity strength
$(1-\xi^2)^2/(1+\xi^2)^2$, at which the impurity band can not be
distinguished from the continuum, and the result is shown in
Fig.\ref{ni}.
Here we have denoted $\xi=\frac{JS}{2}\pi N_F$ as the effective
impurity strength and $\frac{1}{\tau_s
M}=n_i\frac{JS}{M}\frac{\xi}{(1+\xi^2)^2}$ as the effective
concentration. However we suggest that experimentally the actual
concentration should be much lower than the critical values, so that
not only its influence on the exchange coupling $J$ is negligible,
\cite{Novik2005} but also the impurities can be considered isolated
and their coupling, such as RKKY interactions, can be ignored.
Furthermore, we speculate that our results should be still valid
even within a complete quantum treatment of the magnetic impurity.
\cite{ZM}

Interestingly, the existence of LES in the bulk energy gap of a
topological insulator is not special in two dimensions, but is also
true for 3D strong topological insulators (STI). As an example, we
consider the strained HgTe which is believed to be a STI.
\cite{Qi2008a,Dai2008} The effect of magnetic impurities on the
surface states of strained HgTe has been discussed by one of the
authors, \cite{liu2008a} here we focus on its effect in bulk. The
model Hamiltonian describing strained HgTe with time-reversal as
well as inversion symmetries takes the form \cite{Qi2008a,Dai2008}
\begin{equation}
H_{\text{3D}}(\vec{k})=\mathcal{M}_{k}\Gamma ^{1}+A_{\bot
}k_{x}\Gamma ^{5}+A_{\bot }k_{y}\Gamma ^{2}+A_{\Vert }k_{z}\Gamma
^{3} \label{H3D}
\end{equation}
where $\mathcal{M}_{k}=M_{0}-M_{1}(k_x^2+k_y^2)-M_{2}k_{z}^{2}$, and
the representation for Gamma matrices is chosen in such a way that
they are invariant under the joint transformations of inversion and
time-reversal symmetries. \cite{note} Two features are worth
noticing about this model Hamiltonian. Firstly, by setting
$A_{\Vert}=0$, Eq.(\ref{H3D}) recovers the 2D HgTe/CdTe QW model. As
we have discussed in the above, there are always LES in the inverted
regime for arbitrarily strong impurity strength. Secondly, by
comparing Eq.(\ref{H3D}) with the Kane model \cite{winkler} in the
basis of
$\left\vert E,\frac{1}{2}\right\rangle ,$ $\left\vert LH,-\frac{1}{2}%
\right\rangle ,$ $\left\vert E,-\frac{1}{2}\right\rangle ,$ $\left\vert LH,%
\frac{1}{2}\right\rangle $, we observe that they have exactly the
same form. Therefore the matrix form for Kondo-like sp-d exchange
term, \cite{winkler,Novik2005,lcx2008} $H_{ex}\left( \vec{r}\right)
=-\sum_{m}J(\vec{r}-\vec{R}_m) \vec{S}_{m}\cdot \vec{\sigma}$, in
the same basis can be similarly extracted as
\begin{eqnarray}
H_{ex} &=&\left(
\begin{array}{cccc}
-\Delta _{s}e_{z} & 0 & -\Delta _{s}e_{-} & 0 \\
0 & \frac{1}{3}\Delta _{p}e_{z} & 0 & -\frac{2}{3}\Delta _{p}e_{+} \\
-\Delta _{s}e_{+} & 0 & \Delta _{s}e_{z} & 0 \\
0 & -\frac{2}{3}\Delta _{p}e_{-} & 0 & -\frac{1}{3}\Delta _{p}e_{z}%
\end{array}
\right)
\label{Hex}
\end{eqnarray}
where $\Delta_s=yN_{0}J_{s}\langle S\rangle $,
$\Delta_p=yN_{0}J_{p}\langle S\rangle $, \cite{Novik2005} and
$\vec{e}$ is the unit vector along the direction of impurity spin.

Following the methods developed by Fu {\it et. al} on 3D topological
insulators with inversion symmetries, \cite{Fu2007,Liang2007b} the
STI phase characterized by odd number of Dirac points $(k_x,k_y)$ in
the 2D surface states of Hamiltonian (\ref{H3D}) can be analyzed
through two parameters $r=M_{2}/M_{1}$ and $r_{1}=M_{0}/4M_{1}$.
Though we wouldn't elaborate the results in general, some specific
examples are listed below. On a square lattice, for $r=1.5$, there
is one Dirac point at $(0,0)$ for $r_1=0.75$; three Dirac points at
$(0,0),\;(0,\pi)$ and $(\pi,0)$ for $r_1=1.25$; the three Dirac
points then move to $(0,\pi),\; (\pi,0)$ and $(\pi,\pi)$ for
$r_1=2.25$; while for $r_1=3$ there is only one Dirac point again at
$(\pi,\pi)$. For even larger $r_1$ the system evolves out of the STI
phase.

Using the same formulation, \cite{Shiba1968} the full GF for the
system $H=\sum_{k}c_{k}^{\dag }H_{3D}(\vec{k})
c_{k}+\sum_{kk^{\prime }}c_{k}^{\dag }H_{ex}c_{k^{\prime }}$ is
obtained by finding the t-matrix as
\begin{equation}
t_{3D}\left( \omega \right)=\frac{H_{ex}}{1-H_{ex}F\left(
\omega \right)}
\end{equation}
where again the results in Eqs.(\ref{Fmatrix}) and (\ref{Ffunction})
are recovered [$\epsilon_k=0$ automatically here since there are no
kinetic term in the Dirac Hamiltonian (\ref{H3D})] with
$\mathcal{M}_{k}=M_{0}-2M_{1}\left( 2-\cos k_{x}-\cos
k_{y}\right)-2M_{2}\left( 1-\cos k_{z}\right)$ and $D_{k}=\omega
^{2}-\left[ \mathcal{M}_{k}^{2}+2A_{\bot }^{2}\left( 2-\cos
k_{x}-\cos k_{y}\right) +2A_{\shortparallel }^{2}\left( 1-\cos
k_{z}\right)\right]$ in the 3D case. The resonant conditions for LES
are obtained similarly by finding the poles of the full GF in the
bulk energy gap as
\begin{equation}
\Delta_sF_e(\omega)=\pm 1, \;\;
\frac{\Delta_p\sqrt{4-3e^2_z}}{3}F_h(\omega)=\pm 1
\label{resonant3D}
\end{equation}
By numerically plotting $F_{e(h)}$ as a function of $\omega$ in the
bulk energy gap, similar behavior as shown in Fig.\ref{F2D} is found
respectively for STI and non-STI phases using the values of $r$ and
$r_1$ illustrated above. We see that the same conclusion applies to
3D topological insulators. In the STI phase there are always LES in
the bulk energy gap for arbitrary exchange interaction strength,
while when out of STI phase, the LES exist only for very weak
exchange interaction strength. Therefore we believe that the
existence of nonvanishing LES in the bulk energy gap for arbitrary
impurity strength plays the role of a general characterization for
topological insulators.

Experimentally we suggest to detect this effect in the recently
achieved HgTe/CdTe QWs \cite{Bernevig2006D,Konig2007} by doping a
small concentration of Mn$^{+2}$ ions in bulk. Since the LES evolve
with the combination of exchange coupling strength and the magnetic
moments (at the mean-field level), though it is hard to adjust the
impurity exchange coupling strength, it may possible to tune the Mn
moments by a small magnetic field. \cite{Novik2005,lcx2008}. When
the Mn moments are larger than some critical value, there will be
four peaks in the DOS spectrum in STM measurements for the QW width
$d>d_c$, which persist for even larger polarization. When $d<d_c$,
the peaks in the DOS spectrum will be broadened and vanish as the
increasing of the polarization. However, for 3D STI systems with
impurities doped deep in bulk, ARPES measurements will be more
appropriate. We suggest to use this distinct signal to differentiate
experimentally the topological and the conventional insulating
phases.

The authors thank Prof. Shou-Cheng Zhang and Dr. Chao-Xing Liu for
many illuminating discussions.  Q. Liu acknowledges the support of
China Scholarship Council for support. This work is supported by
HKSAR RGC Project No. CUHK 401806.


\begin{thebibliography}{99}
\bibitem{note} Explicitly, we take the
time-reversal operator as $\mathcal{T}=\sigma^{0}\otimes
i\tau^{2}K$, where $K$ is the complex conjugation, and the inversion
operator $\hat{P}=\sigma^{z}\otimes\tau^{0}$. Under these
definitions, the Gamma matrices which are invariant under the joint
transformations of inversion and time-reversal symmetries are
$\Gamma ^{1}=\sigma^{3}\otimes \tau^{0},\Gamma
^{2}=\sigma^{2}\otimes\tau^{0},\Gamma ^{3,4,5}=\sigma^{1}\otimes
\tau^{1,2,3}$.
\end{thebibliography}
\end{document}